\documentclass[aps,a4paper,showpacs,amsfonts,amsmath,amssymb]{ revtex4}

\usepackage{graphicx}
\usepackage{bm}
\usepackage{amsmath,epsfig}
\usepackage{natbib}
\usepackage{dcolumn}
\usepackage[hang,small,bf,centerlast]{caption}

\begin{document}

\title{Economic sector identification in a set of stocks traded at the New York Stock Exchange: a comparative analysis}

\author{C. Coronnello$^{\dagger \star}$, M. Tumminello$^{\dagger}$, 
        F. Lillo$^{\dagger \star \times}$, S. Miccich\`e$^{\dagger \star}$, 
        R.N. Mantegna$^{\dagger \star}$}
\address{$^\dagger$ Dipartimento di Fisica e Tecnologie Relative, Universit\`a degli Studi di Palermo, Viale delle Scienze, Edificio 18, I-90128, Palermo, Italy}
\address{$^\star$ INFM-CNR, Unit\`a di Palermo, Palermo, Italy}
\address{$^\times$ Santa Fe Institute, 1399 Hyde Park Road, Santa Fe, NM 87501, USA}

\begin{abstract}
We review some methods recently used in the literature to detect the existence of a certain degree of common behavior of stock returns belonging to the same economic sector. Specifically, we discuss methods based on random matrix theory and hierarchical clustering techniques. We apply these methods to a set of stocks traded at the New York Stock Exchange. The investigated time series are recorded at a daily time horizon. 

All the considered methods are able to detect economic information and the presence of clusters characterized by the economic sector of stocks. However, different methodologies provide different information about the considered set. Our comparative analysis suggests that the application of just a single method could not be able to extract all the economic information present in the correlation coefficient matrix of a set of stocks.
\end{abstract}

  
\date{\today}

\maketitle

\section{Introduction}

The correlation matrix is one of the major tools useful to study the collective behaviour of systems characterized by the presence of a large number of elements interacting with each other in a given period of time. Multivariate time series are detected and recorded both in experiments and in the monitoring of a wide number of physical, biological and  economic systems. The study of the properties of the correlation matrix has a direct relevance in the investigation of mesoscopic physical systems \cite{Forrester94}, high energy physics \cite{Demasure2003}, information theory and communication \cite{Moustakas2000,Tworzydlo2002,Skipetrov2003}, investigation of microarray data in biological systems \cite{maritan,brown,banavar} and econophysics \cite{Laloux1999,Plerou1999,Mantegna1999,maslov,pafka,sornette,burda}.

In the case of multivariate stock return time series, the correlation matrix may contain information about the economic sectors of the considered stocks \cite{Mantegna1999,Kullmann2000,Bonanno2000,Gopikrishnan2001,Bonanno2001,Kullmann2002,Onnela2002,Giada2002,Bonanno2003,Bonanno2004,Tumminello2005,coronnello}. Recent empirical and theoretical analysis have shown that this information can be detected by using a variety of methods. In this paper we review some of these methods based on Random Matrix Theory (RMT) \cite{Gopikrishnan2001}, correlation based clustering \cite{Mantegna1999}, and topological properties of correlation based graphs \cite{Tumminello2005}. 

In this paper we perform a comparative analysis of the different techniques able to detect economic information out of the correlation matrix. In particular, our attention will be focused on the robustness of these methodologies with respect to statistical uncertainty and their capability in providing economic information. The review opens up with the discussion of the concepts of RMT. As explained below, this methodology can be used to select the eigenvalues and eigenvectors of the correlation matrix less affected by statistical uncertainty. Then we consider clustering techniques. In particular, we focus our attention on two correlation based hierarchical clustering procedures that can be used to obtain a reduced number of similarity measures representative of the whole original correlation matrix. These clustering procedures allow to obtain a number of similarity measures of the order of $n$, when starting from a similarity matrix with $n(n-1)/2$ distinct elements. The first clustering procedure we consider here is the single linkage clustering method that has been repeatedly used to detect a hierarchical organization of stocks. One advantage of using such correlation based clustering procedure is that it provides in a unique way a Minimum Spanning Tree (MST). The second clustering procedure is the average linkage which provides a different taxonomy and the last one is the PMFG. We also consider a recently introduced graph, the Planar Maximally Filtered Graph (PMFG) which extends the number of similarity measures associated to the graph with respect to the ones of the MST \cite{Tumminello2005}, although it has associated the same hierarchical tree.

The present investigation regards the transactions occured at the New York Stock Exchange (NYSE) in the year 2002 for a set of $n=100$ highly capitalized stocks. The analysis is performed on the price returns computed at a daily time horizon. 

The paper is organized as follows: in Section 2 we present the data set used in our empirical investigation. In Section 3 and Section 4 we illustrate the methods used to extract economic information from a correlation matrix of a stock set by using concepts and tools of RMT and hierarchical clustering. In Section 5 we illustrate a recently introduced technique which allows the construction of a graph, the Planar Maximally Filtered Graph (PMFG), obtained by imposing the topological constraint of planarity at each step of its construction. In Section 6 we draw our conclusions.

\section{The Data Set} \label{dataset}

We investigate the statistical properties of price returns for $n=100$ highly capitalized stocks traded at NYSE. In particular, we consider transactions occurred in year 2002. The empirical data are taken from the ``Trades and Quotes'' database, issued by the NYSE.

For each stock and for each trading day we consider the closure price. We compute daily returns as the difference of the logarithms of the closure prices of each successive trading day. To each of the 100 selected stocks an economic sector of activity is associated according to the classification scheme used in the web--site {\tt{http://finance.yahoo.com}}. The economic sectors are reported in Table \ref{classification}, together with the number of stocks belonging to each of them (third column). 
\begin{table}
\begin{center}
\caption{Economic sectors of activity for 100 highly capitalized stocks traded at NYSE. The classification is done according to the methodology used in the web--site {\tt{http://finance.yahoo.com}}. The second column contains the economic sector and the third column contains the number of stocks belonging to the sector.} \label{classification}
\vspace{.5 cm}
\begin{tabular}{||c|l|c||}
\hline
                     & ${\rm{SECTOR}}$      ~&~{\rm{NUMBER}}~\cr \hline
                  1  & Technology            & $ 8$          \\
                  2  & Financial             & $24$          \\
                  3  & Energy                & $ 3$          \\
                  4  & Consumer non-Cyclical & $11$          \\
                  5  & Consumer Cyclical     & $ 2$          \\
                  6  & Healthcare            & $12$          \\
                  7  & Basic Materials       & $ 6$          \\
                  8  & Services              & $20$          \\
                  9  & Utilities             & $ 2$          \\
                 10  & Capital Goods         & $ 6$          \\
                 11  & Transportation        & $ 2$          \\
                 12  & Conglomerates         & $ 4$          \cr \hline
\end{tabular}
\end{center}
\end{table}

The correlation coefficient between two stock return time series is defined as
\begin{eqnarray}
                \rho_{ij}=\frac{\langle r_i r_j\rangle -\langle r_i\rangle\langle r_j\rangle}
{\sqrt{(\langle r_i^2\rangle-\langle r_i\rangle^2)(\langle r_j^2\rangle -\langle r_j\rangle^2)}} 
                \qquad \qquad i,j=1, \dots, n
\end{eqnarray} 
where $n$ is the number of stocks, $i$ and $j$ label the stocks, $r_i$ is the logarithmic return defined by $r_i = \ln P_i (t) - \ln P_i (t - \Delta t)$, $P_i (t)$ is the value of the stock price $i$ at the trading time $t$ and $\Delta t$ is the time horizon at which one computes the returns. In this work the correlation coefficient is computed between synchronous return time series. The correlation coefficient matrix is an $n \times n$ matrix whose elements are the correlation coefficients $\rho_{ij}$. An euclidean metric distance between pair of stocks can be rigorously determined \cite{Gower1966} by defining
\begin{eqnarray}
                d_{ij}=\sqrt{2 (1-\rho_{ij})} \label{distance}
\end{eqnarray} 
With this choice $d_{ij}$ fulfills the three axioms of a metric ­ (i) $d_{ij} = 0$ if and only if $i = j$ ; (ii) $d_{ij} = d_{ji}$ and (iii) $d_{ij} \le d_{ik} + d_{kj}$. 

\section{Random Matrix Theory} \label{RMT}

Random Matrix Theory \cite{Metha90} was originally developed in nuclear physics and then applied to many different fields. In the context of asset set management RMT is useful because it allows to quantify the statistical uncertainty, due to the finiteness of the time series under investigation, in the estimation of the correlation matrix. 
Let us consider $n$ assets whose returns are described by $n$ time series of length $T$. Let us also suppose that such returns are independent Gaussian random variables with zero mean and variance $\sigma^2$. In the limit $T\to \infty$, the correlation matrix of this set of variables is simply the identity matrix. However, when $T$ is finite the correlation matrix will in general be different from the identity matrix. RMT allows to prove that for large values of $T$ and $n$, with a fixed ratio $Q=T/n \geq 1$, the eigenvalue spectral density of the covariance matrix is given by
\begin{equation}
                \rho(\lambda)=\frac{Q}{2\pi\sigma^2\lambda}
                                       \sqrt{(\lambda_{max}-\lambda)
                                             (\lambda-\lambda_{min})},  \label{zerofactor}
\end{equation}
where $\lambda_{min}^{max}=\sigma^2 (1+1/Q\pm 2\sqrt{1/Q})$. The function $\rho(\lambda)$ is defined as the probability density function of eigenvalues. The above spectral density is different from zero in the interval $]\lambda_{min},\lambda_{max}[$. In the case of a correlation matrix one can set $\sigma^2=1$. It is evident that the spectrum described by Eq.~\ref{zerofactor} is different from $\delta(\lambda-1)$, i.e. from an identity correlation matrix. RMT is therefore a powerful tool able to quantify the role of the finiteness of the time series length on the spectral properties of the correlation matrix. 

RMT has been applied to the investigation of correlation matrices of financial asset returns \cite{Laloux1999,Plerou1999} and it has been shown that the spectrum of a typical set can be divided in three classes of eigenvalues. The largest eigenvalue is totally incompatible with Eq.~\ref{zerofactor} and it is usually thought as describing the common behavior of the stocks composing the set. This fact leads to another working hypothesis according to which the part of correlation matrix which is orthogonal to the eigenvector corresponding to the first eigenvalue has a spectral density $\rho(\lambda)$ described by Eq. \ref{zerofactor}. Under these assumptions, the variance of the part not explained by the highest eigenvalue is given by $\sigma^2=1-\lambda_1/n$ and such value is used to estimate of the variance to be used in Eq.~\ref{zerofactor} to compute $\tilde\lambda_{min}$ and $\tilde\lambda_{max}$. Following these lines, previous studies have shown that a fraction of the order of few percent of the eigenvalues are also incompatible with the RMT because they fall outside the interval $]\tilde\lambda_{min},\tilde\lambda_{max}[$ computed by using the above estimate of $\sigma^2$. It is therefore natural to assume that these are the eigenvalues describing the economic information stored in the correlation matrix. The remaining large part of the eigenvalues is between  $\tilde\lambda_{min}$ and $\tilde\lambda_{max}$ and thus one cannot say whether any information is contained in the corresponding eigenspace. 

The fact that, under certain assumptions, RMT allows to identify the  part of the correlation matrix containing economic information suggested some authors to use RMT for showing that the eigenvectors associated to eigenvalues not explained by RMT describe economic sectors. Specifically, the suggested method \cite{Gopikrishnan2001} is the following. One computes the correlation matrix and finds the spectrum ranking the eigenvalues such that $\lambda_k>\lambda_{k+1}$. The  eigenvector corresponding to $\lambda_k$ is denoted ${\bf u}^k$. The set of investigated stocks is partitioned in $S$ sectors $s=1,2,...,S$ according to their economic activity (for example by using classification codes such as the one of the Standard Industrial Classification code or Forbes). One then defines a $S\times n$ projection matrix ${\bf P}$ with elements $P_{si}=1/n_s$ if stock $i$ belongs to sector $s$ and $P_{si}=0$ otherwise. Here $n_s$ is the number of stocks belonging to sector $s$. For each eigenvector ${\bf u}^k$ one computes
\begin{equation}
                 X^k_s\equiv\sum_{i=1}^n P_{si}[u_i^k]^2 \label{projectioneq}
\end{equation}
This number gives a measure of the role of a given sector $s$ in explaining the composition of eigenvector ${\bf u}^k$. Thus when a given eigenvector has a large value of $X^k_s$ for only one (or few) sector $s$, one can conclude that the eigenvector describes that economic sector (or a linear combination of a few of them). Finally, it is worth noting that the implementation of this method requires the {\it a priori} knowledge of the sector for each stock.

\medskip

The above concepts and tools can be used to infer economic information about the set of stocks considered in Section \ref{dataset}. For a time horizon of one trading day the largest eigenvalue is $\lambda_1=39.2$ clearly incompatible with RMT and suggesting a driving factor common to all the stocks. This is usually interpreted to be the ``market mode" as described in widespread market models, such as the Capital Asset Pricing Model. The analysis of the components of the corresponding eigenvector confirms this interpretation. In fact the mean component of the first eigenvector is $0.096$ and the standard deviation is $0.027$ showing that all the stocks contribute in a similar way to the eigenvector ${\bf u}^1$.

In our data $Q=T/n=2.5$ and the threshold value $\lambda_{max}$ without taking into account the first eigenvalue is $\lambda_{max}=2.66$. This implies that RMT considers as signal only the first three eigenvalues $\lambda_1$, $\lambda_2=5.29$, and $\lambda_3=2.85$. On the other hand if we remove the contribution of the first eigenvalue with the procedure discussed in section \ref{RMT} we get $\tilde\lambda_{max}=1.62$, indicating that the first $9$ eigenvalues could contain economic information. This result shows the importance of taking into account the role of the first eigenvalue.

\begin{figure}[ptb]
\begin{center}
              \includegraphics[scale=.45]{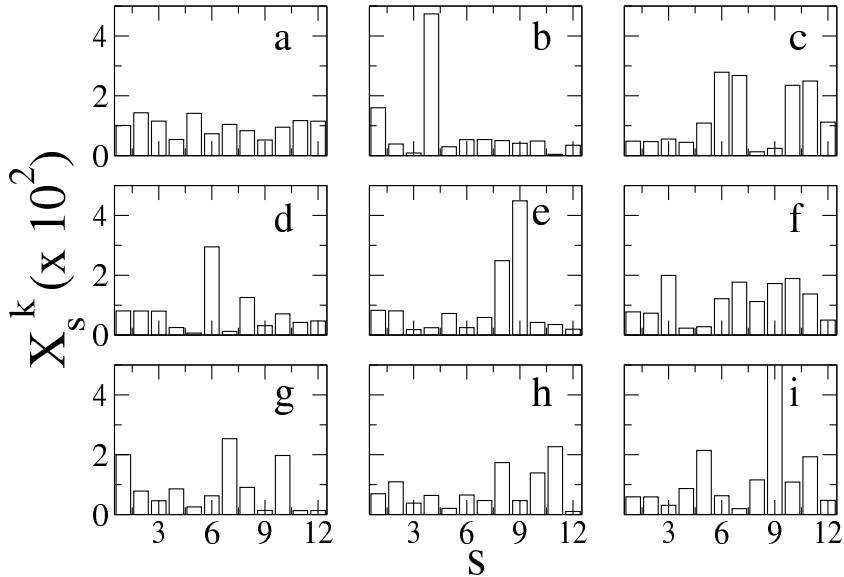}
               \caption{$X^k_s$ of Eq.~\ref{projectioneq} for the first nine eigenvectors of the correlation matrix of daily returns of $100$ NYSE stocks. The order of sectors is the same as in Table \ref{classification}.} \label{projection1day}
\end{center}
\end{figure}

Figure~\ref{projection1day} shows $X^k_s$ of Eq.~\ref{projectioneq} of the first $9$ eigenvalues. Panel (a) shows that all the sectors contribute roughly in a similar way to the first eigenvector. On the other hand eigenvectors $2$, $4$, $5$, and $9$ are characterized by one prominent sector. Specifically, the second eigenvector shows a large contribution from the sector Consumer non-Cyclical ($s=4$), the fourth eigenvector has a significant contribution from the Healthcare sector ($s=6$), the fifth eigenvector is dominated by Utilities ($s=9$) and Services ($s=8$) and the ninth eigenvector has also a large peak for the sector Utilities ($s=9$).  The other eigenvectors do not show significant peaks, indicating that probably either these are eigenvectors mixing different groups or they are strongly affected by statistical uncertainty (``noise dressed'').

\section{Hierarchical Clustering Methods}

Another approach used to detect the information associated to the correlation matrix is given by the correlation based
hierarchical clustering analysis. Consider a set of $n$ objects and suppose that a similarity measure, e.g. the correlation coefficient, between pairs of elements is defined. Similarity measures can be written in a $n\times n$ similarity matrix. The hierarchical clustering methods allow to hierarchically organize the elements in clusters. The result of the procedure is a rooted tree or dendrogram giving a quantitative description of the clusters thus obtained. 

It is worth noting that hierarchical clustering methods can as well be applied to distance matrices. 

A large number of hierarchical clustering procedures can be found in the literature. For a review about the classical techniques see for instance Ref. \cite{Anderberg}. In this paper we focus our attention on the Single Linkage Cluster Analysis (SLCA), which was introduced in finance in Ref. \cite{Mantegna1999} and the Average Linkage Cluster Analysis (ALCA).

\subsection{Single Linkage Correlation Based Clustering} \label{SLCA}

The Single Linkage Cluster Analysis is a filtering procedure based on the estimation of the subdominant ultrametric distance \cite{Rammal1986} associated with a metric distance obtained from the correlation coefficient matrix of a set of $n$ stocks. This procedure, already used in other fields, allows to extract a hierarchical tree from a correlation coefficient
matrix by means of a well defined algorithm known as nearest neighbor single linkage clustering algorithm \cite{Mardia}. The methodology also allows to associate a MST to the hierarchical tree. Such association is essentially unique. By using this technique, it is therefore possible to reveal both topological (through the MST) and taxonomic (through the hierarchical tree) aspects of the correlation among stocks.

In the SLCA algorithm, at each step, when two elements or one element and a cluster or two clusters $p$ and $q$ merge in a wider single cluster $t$, the distance $d_{tr}$ between the new cluster $t$ and any cluster $r$ is recursively determined as follows:  
\begin{equation}
                d_{tr}=\min \{ d_{pr},d_{qr}\}
\end{equation}
thus indicating that the distance between any element of cluster $t$ and any element of cluster $r$ is the shortest distance between any two entities in clusters $t$ and $r$. By applying iteratively this procedure $n-1$ of the $n(n-1)/2$ distinct elements of the original correlation coefficient matrix are selected. 

The distance matrix obtained by applying the SLCA is an ultrametric matrix comprising $n-1$ distinct selected elements. The subdominant ultrametric is the ultrametric structure closest to the original metric structure \cite{Rammal1986}. The ultrametric distance $d_{ij}^<$ between element $i$ belonging to cluster $t$ and element $j$ belonging to cluster $r$ is defined as the distance between clusters $t$ and $r$. Ultrametric distances $d_{ij}^<$ are distances satisfying the inequality $d^<_{ij} \le  \max \{d^<_{jk},d^<_{kj}\}$ stronger than the customary triangular inequality $d_{ij} \le  d_{ik} + d_{kj}$ \cite{Rammal1986}. The SLCA has associated a correlation matrix which is associated to the subdominant ultrametric distance matrix obtained from the original correlation coefficient matrix. It can be obtained starting from the ultrametric distances $d_{ij}^<$ and making use of Eq. \ref{distance}.

The MST is a graph without loops connecting all the $n$ nodes with the shortest $n - 1$ links amongst all the links between the nodes. The selection of these $n-1$ links is done according to some widespread algorithm \cite{Papadimitriou82} and can be summarized as follows:
\begin{enumerate} 
\item Construct an ordered list of pairs of stocks $L_{ord}$, by ranking all the possible pairs according to their distance $d_{ij}$. The first pair of $L_{ord}$ has the shortest distance.
\item The first pair of $L_{ord}$ gives the first two elements of the MST and the link between them. 
\item The construction of the MST continues by analyzing the list $L_{ord}$. At each successive stage, a pair of elements is selected from $L_{ord}$ and the corresponding link is added to the MST only if no loops are generated in the graph after the link insertion.
\end{enumerate}
Different elements of the list are therefore iteratively included in the MST starting from the first two elements of $L_{ord}$. As a result, one obtains a graph with $n$ vertices and $n-1$ links. For a didactic description of the method used to obtain the MST one can consult Ref. \cite{MS00}

In Ref. \cite{Gower1969} the procedure briefly sketched above has been shown to provide a MST which is associated to the same hierarchical tree of the SLCA. In Ref. \cite{TumminelloManuscript} it is proved that the correlation matrix obtained by the SLCA is always positive definite when all the elements of the obtained correlation matrix are non negative. This condition is rather common in financial data. 

The effectiveness of the SLCA in pointing out the hierarchical structure of the investigated set has been shown by several studies  \cite{Mantegna1999,Bonanno2000,Bonanno2001,Kullmann2002,Onnela2002,Bonanno2003,Bonanno2004,Micciche2003,Dimatteo2004}.

\medskip

\begin{figure}[ptb]
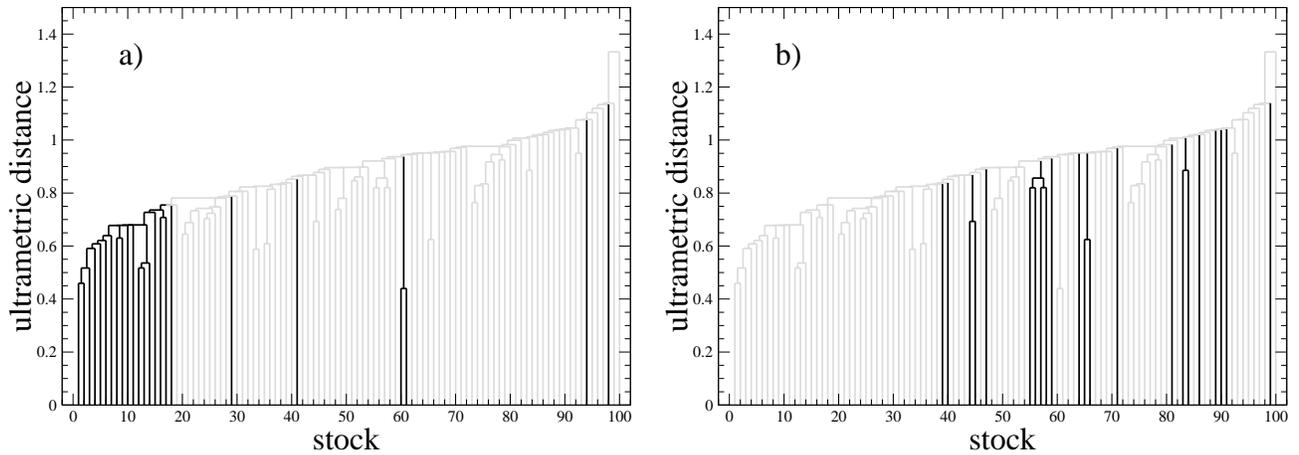

\begin{center}
              \hbox{
              \includegraphics[scale=0.35]{clHTsl_NYSE100_2002_day_financial.eps}
              \hspace{.20 cm}
              \includegraphics[scale=0.35]{clHTsl_NYSE100_2002_day_services.eps}
              }
\end{center}
              \caption{Hierarchical tree obtained by using the SLCA starting from the daily price returns of 100 highly capitalized stocks traded at NYSE. Only transactions occurred in year 2002 are considered. In panel (a) the Financial economic sector is highlighted. In panel (b) the Services economic sector is highlighted.} \label{HT_sl_day}
\end{figure}

\begin{figure}[ptb]
\begin{center}
              \includegraphics[scale=0.47]{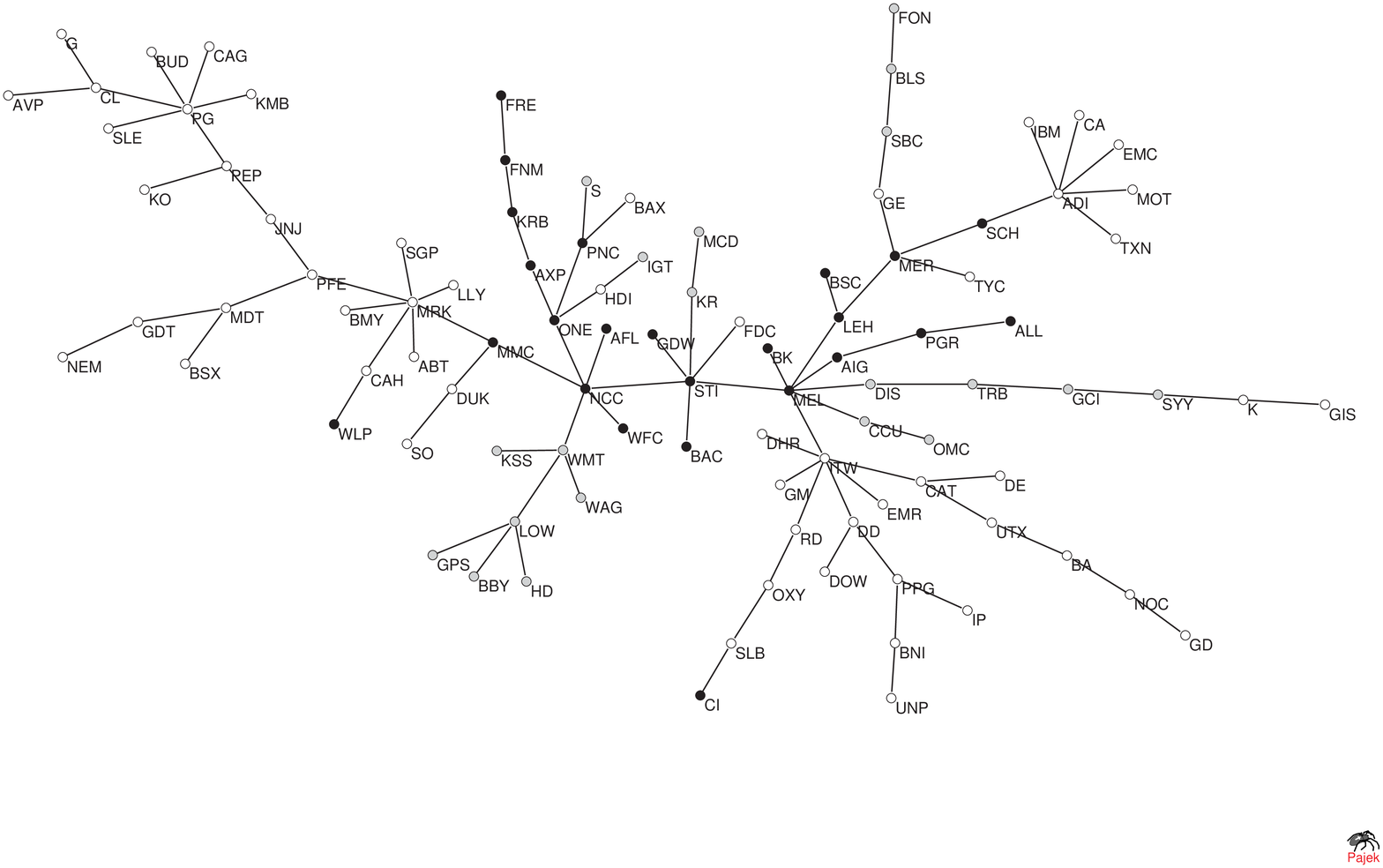}
              \caption{MST obtained starting from the daily price returns of 100 highly capitalized stocks traded at NYSE. Only transactions occurred in year 2002 are considered. The Financial economic sector (black) and the Services (gray) economic sector are highlighted.} \label{MST_sl_day}
\end{center}
\end{figure}

The methodology sketched above can again be applied to the set of stocks considered in Section \ref{dataset}. The results obtained by using the SLCA for the daily returns are summarized in Fig. \ref{HT_sl_day} and Fig. \ref{MST_sl_day} that show the hierarchical tree and the MST, respectively.

The hierarchical tree shows that there exists a significant level of correlation in the market, and also in many cases clustering can be observed. For example, the $60_{th}$ and $61_{st}$ stocks of Fig. \ref{HT_sl_day}, {\em{Freddie Mac}} (FRE) and {\em{Fannie Mae}} (FNM), belonging to the Financial sector, are linked together at an ultrametric distance $d^<\approx0.45$ corresponding to a correlation coefficient as high as $\rho=0.90$. We focus here our attention on the two sectors with the largest number of stocks, which are the Financial sector ($s=2$) and the Services sector ($s=8$). Panel (a) of Fig. \ref{HT_sl_day}, where the stocks of the Financial sector are highlighted, gives an example in which most of the stocks belonging to the same economic sector are clustered together. In fact, a cluster including 18 stocks from position 1 to position 18 can be observed. The remaining six stocks of the Financial sector are {\em{All State Corp}} (ALL) at position 29, {\em{Aflac}} (AFL) at position 41, FRE at position 60, FNM at position 61, {\em{Well Point}} (WLP) at position 94 and {\em{CIGNA}} (CI) at position 98. ALL and AFL are insurance companies, FNM and FRE are mortgage companies while WLP and CI provide health or commercial benefits for the employees of other companies. On the contrary, all the 18 stocks that cluster together are relative to banks, financial institutions or companies that provide financial services, with the only exception of {\em{Progressive Corp}} (PGR) at position 16 which is an insurance company. This result, obtained for the Financial sector, is representative of the capability of SLCA to evidentiate economic subsectors. In panel (b) of Fig. \ref{HT_sl_day} the stocks of the Services sector are highlighted. The Services sector gives an example of the case in which stocks belonging to the same economic sector are poorly clustered. In fact, the biggest cluster is composed of only four stocks. Specifically, the cluster is formed by {\em{Lowes Companies}} (LOW), {\em{Home Depot}} (HD), {\em{Kohls Corporation}} (KSS) and {\em{Wal-Mart Stores}} (WMT) located from position 55 to position 58 in the figure. It is worth noting that all of them are classified as retail companies at the level of economic subsector. This results suggests that economic subsectors are highly characterizing for Services companies. Another economic sector that shows a significant level of clustering is the Basic Materials sector. The smallest economic sectors, i.e. Energy, Consumer Cyclical and Utilities are completely clustered. The stocks of the Technology, Consumer Non Cyclical and Healthcare economic sectors are not fully clustered. It is worth noting that the results regarding the Consumer Non Cyclical and the Healthcare economic sectors are different from those obtained by the RMT.

\begin{figure}[ptb]
\begin{center}
              \includegraphics[scale=0.47]{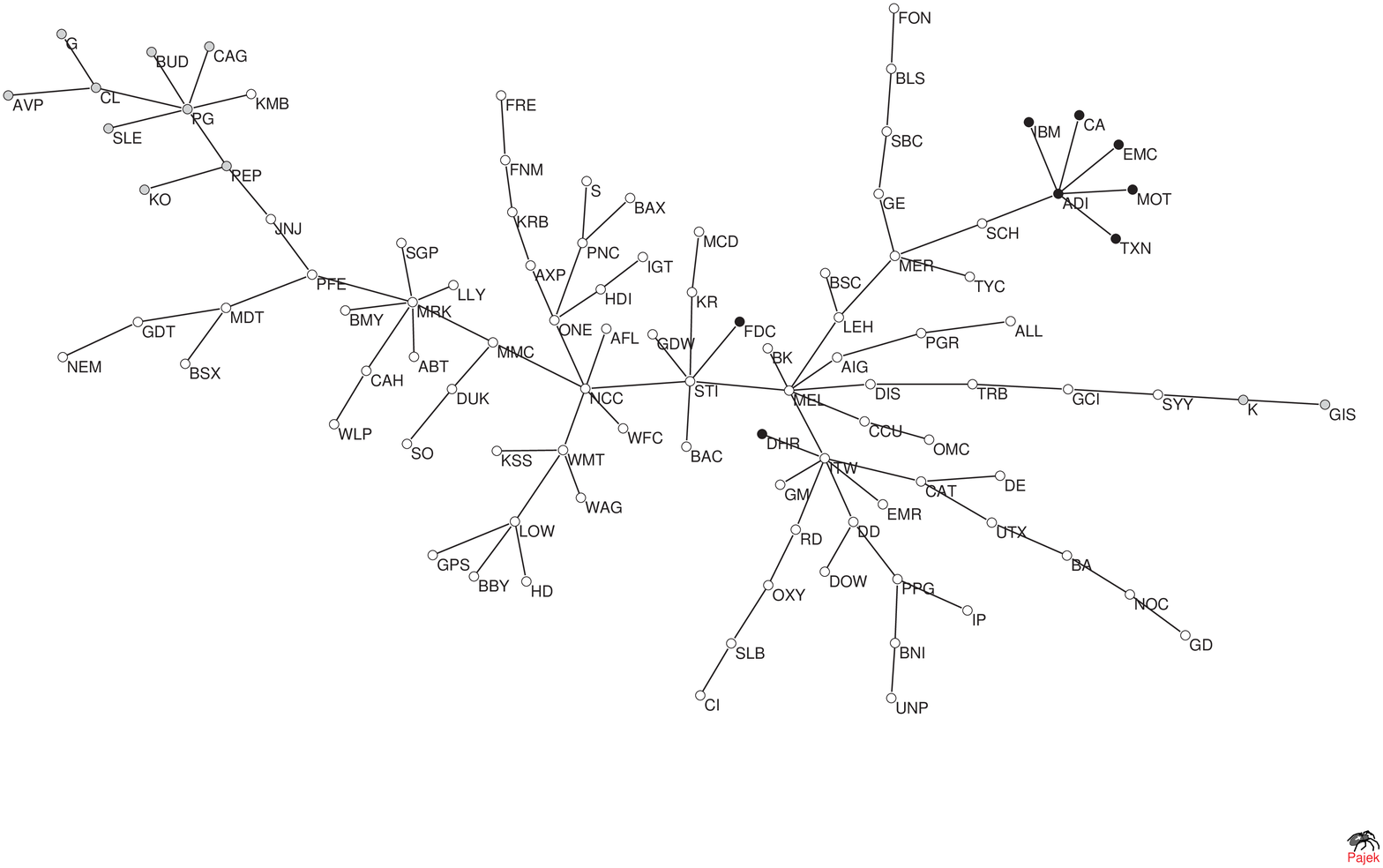}
              \caption{MST obtained starting from the daily price returns of 100 highly capitalized stocks traded at NYSE. Only transactions occurred in year 2002 are considered. The Technology economic sector (black) and the Consumer Non Cyclical (gray) economic sector are highlighted.} \label{MST_sl_day_TCNC}
\end{center}
\end{figure}

The MST partly confirms the above results. In Fig. \ref{MST_sl_day} the stocks belonging to the Financial economic sector (black circles) and the Services economic sector (gray circles) are indicated and the results agree with those shown by the hierarchical tree. However, the MST gives slightly different informations from the hierarchical tree when considering the Technology, the Consumer Non Cyclical and the Healthcare economic sectors. In order to illustrate this point, in Fig. \ref{MST_sl_day_TCNC} the stocks belonging to the Technology economic sector (black circles) and the Consumer Non Cyclical economic sector (gray circles) are indicated. The figure shows that 6 out of 8 stocks of the Technology sector are gathered around {\em{Analog Devices Inc}} (ADI), while 8 out of 11 stocks of the Consumer Non Cyclical sector are gathered around {\em{Procter $\&$ Gamble}} (PG). This observation suggests that the MST is carrying information which is additional to the one present in the hierarchical tree. 

Finally, the topology of the MST shows the existence of several stocks with relatively large degree (number of links with other elements). This is incompatible with the hypothesis that the set of stocks is described by a single factor model \cite{Bonanno2003}.

\subsection{Average Linkage Correlation Based Clustering} \label{ALCA}

The Average Linkage Cluster Analysis is another hierarchical clustering procedure \cite{Anderberg} that can be described by considering either a similarity or a distance measure. As in the previous case, here we consider the distance matrix ${\bf{D}}$. The following procedure performs the ALCA giving as an output a rooted tree and an ultrametric matrix ${\bf{D}}^<$ of elements $d^<_{ij}$:
\begin{enumerate} 
\item Set ${\bf{T}}$ as the matrix of elements such that ${\bf{T}} = {\bf{D}}$.
\item Select the minimum distance $t_{hk}$ in the distance matrix ${\bf{T}}$. Note that after the first step of construction $h$ and $k$ can be simple elements (i.e. clusters of one element each) or clusters (sets of elements).
\item Merge cluster $h$ and cluster $k$ into a single cluster, say $h$. The merging operation identifies a node in the rooted tree connecting clusters $h$ and $k$ at the distance $t_{hk}$. Furthermore to obtain the ultrametric matrix it is sufficient that $\forall \, i \in h$ and $\forall \, j \in k$ one sets $d^<_{ij}=d^<_{ji}=t_{hk}$. 
\item Redefine the matrix ${\bf{T}}$:
\begin{eqnarray} \label{negspin}
\left \{  \begin{aligned}
        &   t_{hj}= \frac{N_h\,t_{hj}+N_k \, t_{kj}}
                               {N_h+N_k}  & 
                           ~~~~\text{ if } j\neq h \,{\rm{and}} \, j\neq k\\
        &                 \nonumber \\
        &    t_{ij}=t_{ij} & 
                           ~~~~\text{ otherwise, }\\
\end{aligned} \right.
\end{eqnarray}
where $N_h$ and $N_k$ are the number of elements belonging respectively to the cluster $h$ and to the cluster $k$. Note that if the dimension of ${\bf{T}}$ is $m \times m$ then the dimension of the redefined ${\bf{T}}$ is $(m-1) \times (m-1)$ because of the merging of clusters $h$ and $k$ into the cluster $h$.
\item If the dimension of ${\bf{T}}$ is bigger than one then go to step 2 else Stop.   
\end{enumerate}
By replacing point $4$ of the above algorithm with the following item\\

\indent $4.$  Redefine the matrix ${\bf{T}}$:
\begin{equation}\label{negspin2}
\left \{ \begin{aligned}
        &  t_{hj}= Min \left[t_{hj}, t_{kj}\right] & 
                           ~~~~\text{ if } j\neq h \,{\rm{and}} \, j\neq k \nonumber\\
        &  t_{ij}=t_{ij} & 
                           ~~~~\text{ otherwise, }\\
\end{aligned} \right.
\end{equation}
one obtains an algorithm performing the SLCA which is therefore equivalent to the one described in the previous section. The algorithm can be easily adapted for working with similarities instead of distances. It is just enough to exchange the distance matrix ${\bf{D}}$ with a similarity matrix (for instance the correlation matrix) and replace the search for the minimum distance in the matrix ${\bf{T}}$ in point $2$ of the above algorithm with the search for the maximal similarity. 

It is worth noting that the ALCA can produce different hierarchical trees depending on the use of a similarity matrix or a distance matrix. More precisely, different dendrograms can result for the ALCA due to the non linearity of the transformation of Eq. \ref{distance}. This problem does not arise in the SLCA because Eq. \ref{distance} is a monotonic transformation and then ranking the connections with respect to correlations from the greatest to the smallest gives the same results than ranking the connections with respect to distances from the smallest to the greatest.

Unlike the case of the SLCA, it is not possible to uniquely associate a spanning tree to the hierarchical tree obtained by using the ALCA. This is due to the fact that in the ALCA the distance between clusters is defined as the mean distance between the elements of the clusters. Such choice hides the possibility of defining in a unique way what is the path connecting two stocks.

\medskip

\begin{figure}
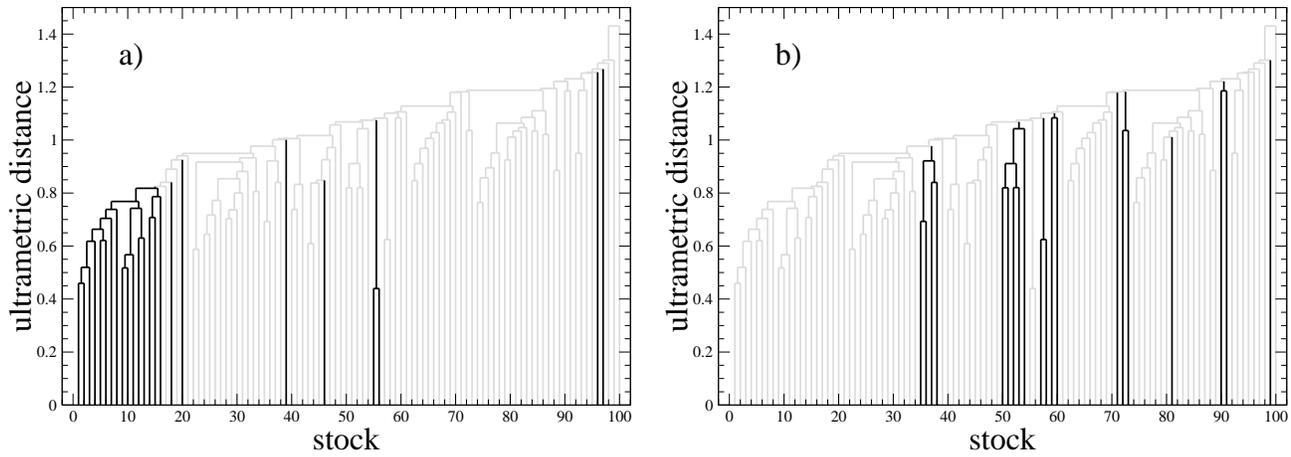
 
\begin{center} 
              \hbox{ 
              \includegraphics[scale=0.35]{clHTal_NYSE100_2002_day_financial.eps} 
              \hspace{.20 cm} 
              \includegraphics[scale=0.35]{clHTal_NYSE100_2002_day_services.eps} 
              } 
              \caption{Dendrogram associated to the ALCA performed on daily returns of a set of 100 stocks traded at NYSE in 2002. Panel (a): The black lines are identifying stocks belonging to the Financial sector. Panel (b): The black lines are identifying stocks belonging to the Services sector} \label{daydendroaverage} 
\end{center} 
\end{figure}

As much as in the case of SLCA we can apply the above methodology to the empirical data described in Section \ref{dataset}. Here we analyze the dendrogram of Fig. \ref{daydendroaverage} obtained by applying the ALCA to the correlation based distance matrix of the daily returns. Once again, to provide representative examples we focus our attention to the two sectors with the largest number of stocks. 

As in Fig. \ref{HT_sl_day}, in panel (a) of Fig. \ref{daydendroaverage} the black lines are identifying stocks of the Financial sector. The figure shows that 16 out of 24 stocks belonging to the Financial sector cluster together. The distance between the elements of this cluster are lower than the average distance. Exceptions are {\em{Golden West Financial Corp}} (GDW) at position 18, AFL at position 20, ALL at position 39, {\em{Charles Schwab Corp}} (SCH) at position 46, FRE at position 55, FNM at position 56, WLP at positions 96 and CI and at positions 97. Interestingly, AFL, ALL, FRE, FNM, WLP and CI are distant from the observed cluster also when considering the SLCA, as shown in panel (a) of Fig. \ref{HT_sl_day}. In panel (b) of Fig. \ref{daydendroaverage} the black lines are identifying the $20$ stocks belonging to the Services sector. Here a cluster of five stocks, from position 50 to position 54 is observed. Moreover, stocks at positions from 50 to 53 are the same ones that can be found in the cluster observed in panel (b) of Fig. \ref{HT_sl_day}, from position 55 to position 58. The fifth stock in the cluster is {\em{WalGreen Company}} (WAG) which is again a company classified as retail at the level of economic subsector. Another cluster of four stocks is observed from position 35 to position 38. Interestingly, the two stocks at positions 35 and 36 belong to the Printing \& Publishing economic subsector, while the two companies at positions 37 and 38 belong to the Broadcasting economic subsector. This result provides a different example which is representative of the capability of hierarchical clustering methods to evidentiate economic subsectors. 

\begin{figure}[ptb]
\begin{center}
              \hbox{
              \includegraphics[scale=0.35]{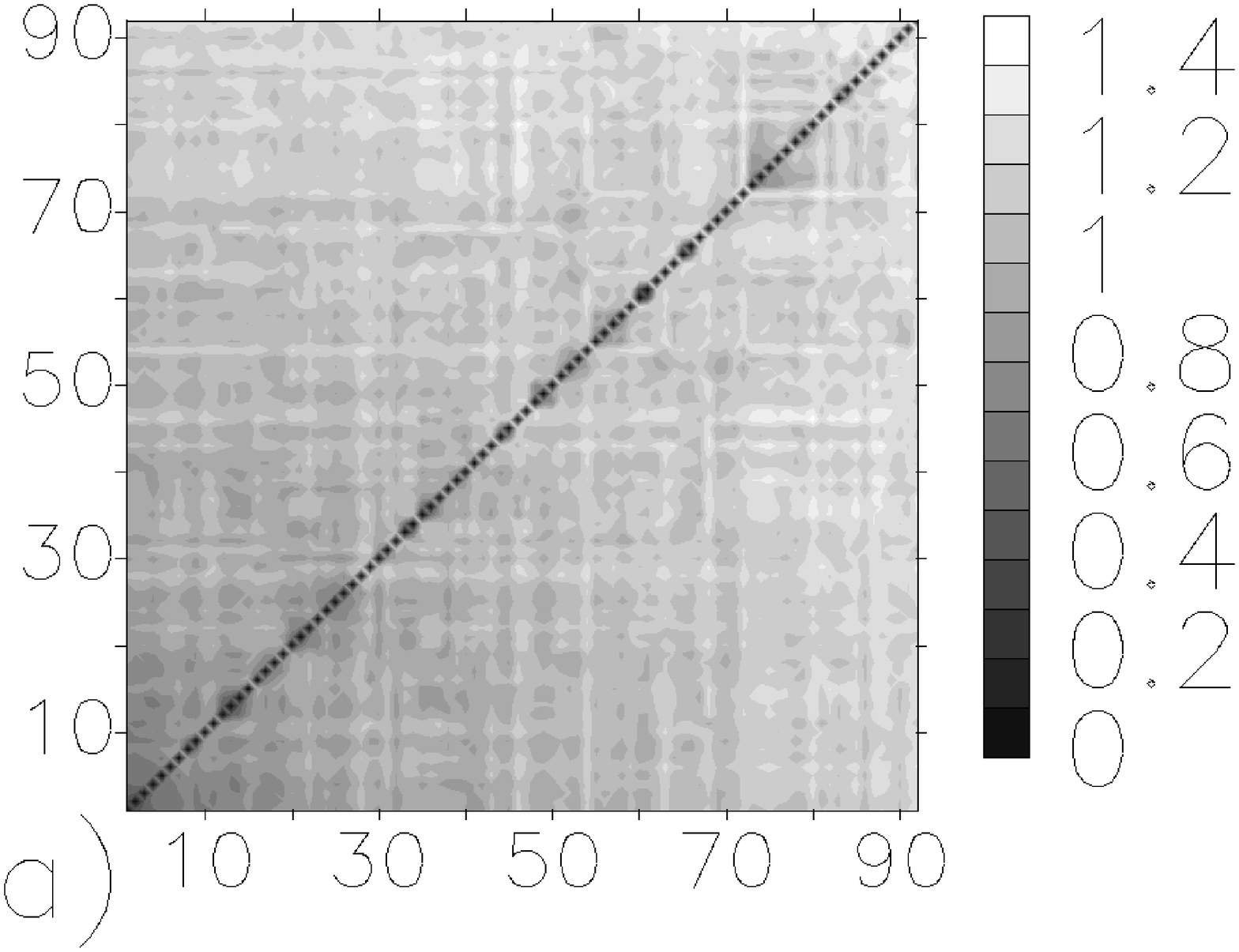}
              \hspace{.20 cm}
              \includegraphics[scale=0.35]{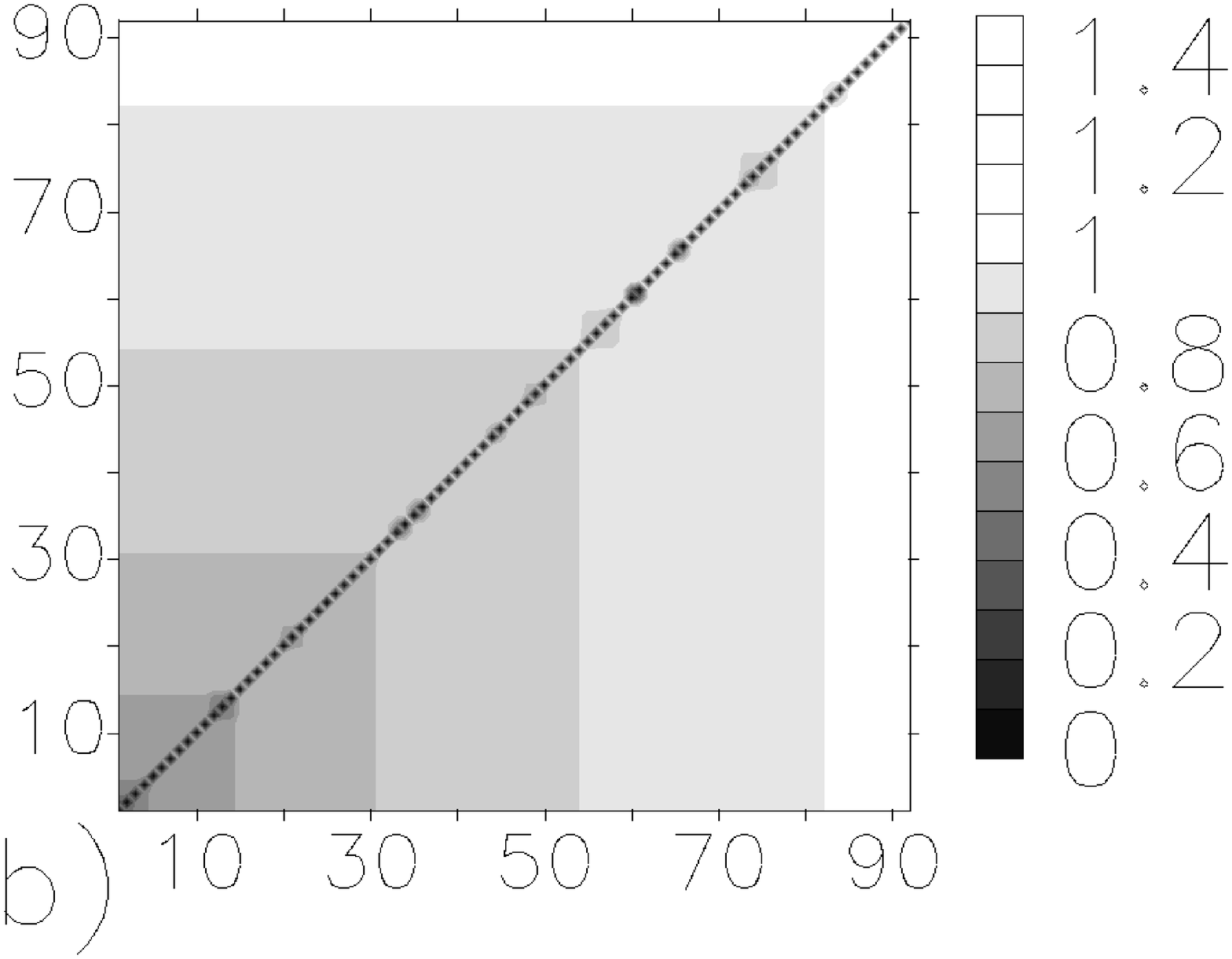}
              }
              \caption{Contour plots of the original distance matrix (panel (a)) and of the one associated to the ultrametric distance (panel (b)) obtained by using the SLCA for the daily price returns of 100 highly capitalized stocks traded at NYSE. Only transactions occurred in year 2002 are considered. Here the stocks are identified by a numerical label ranging from 1 to 100 and ordered according to the hierarchical tree of Fig. \ref{HT_sl_day}. The figure gives a pictorial representation of the amount of information which is filtered out by using the SLCA.} \label{MATRIX_sl_day}
\end{center}
\end{figure}

\begin{figure}[ptb]
\begin{center}
              \hbox{
              \includegraphics[scale=0.35]{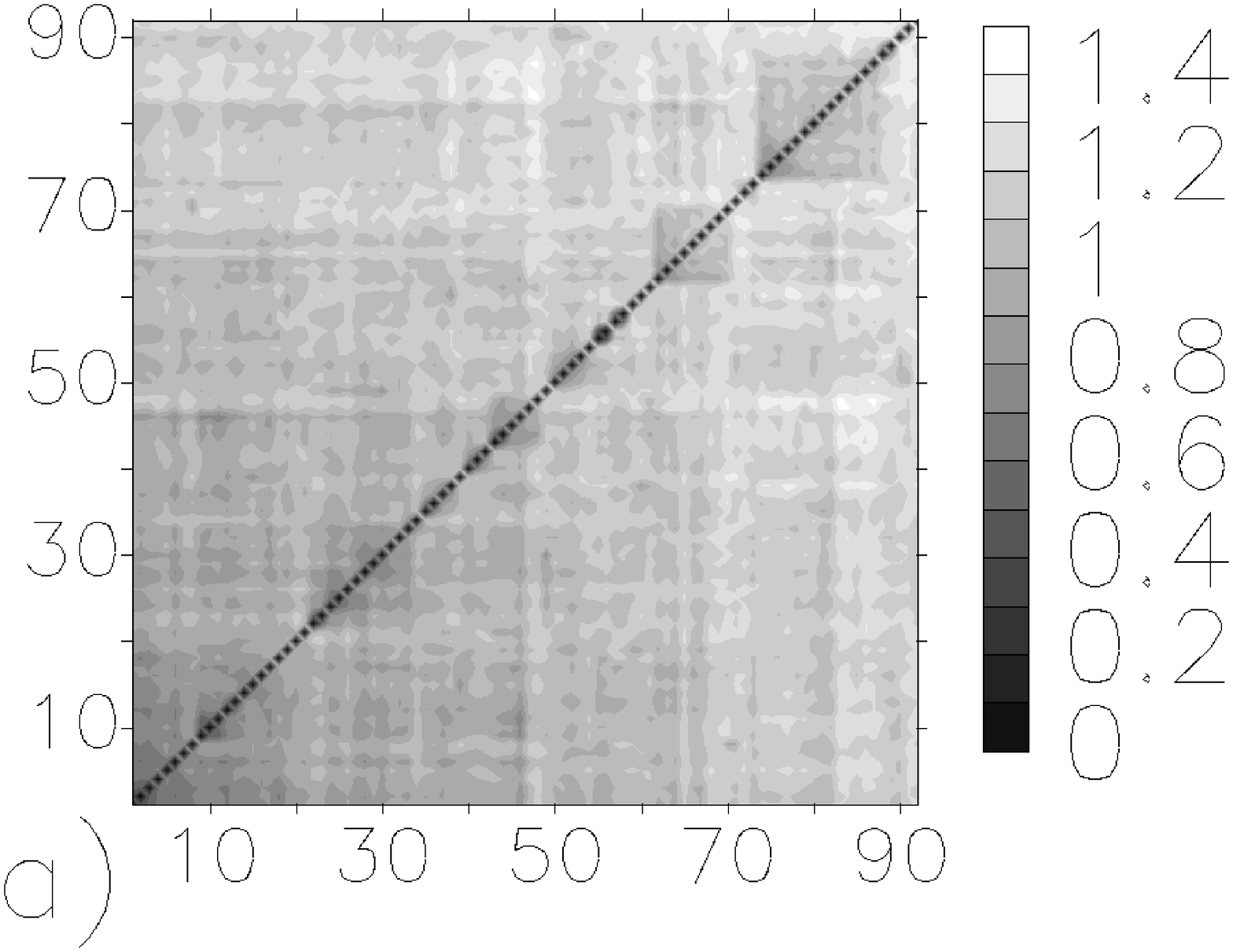}
              \hspace{.20 cm}
              \includegraphics[scale=0.35]{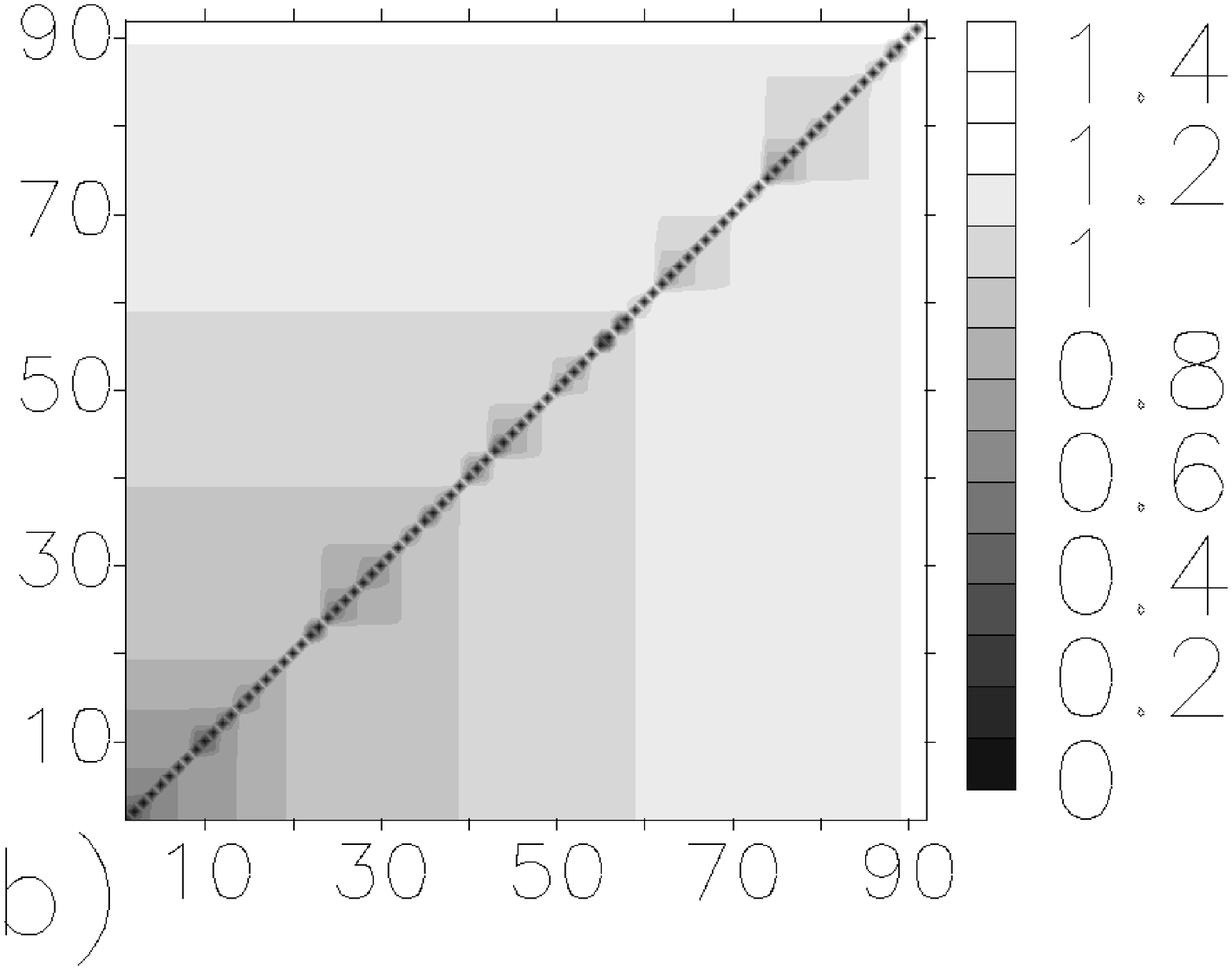}
              }
              \caption{Contour plots of the original distance matrix (panel (a)) and of the one associated to the ultrametric distance (panel (b)) obtained by using the ALCA for the daily price returns of 100 highly capitalized stocks traded at NYSE. Only transactions occurred in year 2002 are considered. Here the stocks are identified by a numerical label ranging from 1 to 100 and ordered according to the hierarchical tree of Fig. \ref{daydendroaverage}. The figure gives a pictorial representation of the amount of information which is filtered out by using the ALCA.} \label{MATRIX_al_day}
\end{center}
\end{figure}

A comparison of the results obtained by using the SLCA and the ALCA shows a substantial agreement between the output of these two methods. However, a refined comparison shows that the ALCA provides a more structured hierarchical tree. In Fig. \ref{MATRIX_sl_day} and Fig. \ref{MATRIX_al_day} we show a graphical representation of the original distance matrix done in terms of a contour plot. In the contour plot the gray scale represents the values of distances among stocks. In the figures we use as stock order the one obtained by SLCA and ALCA respectively. In both cases we also show the associated ultrametric matrices. A direct comparison of the ultrametric distance matrices confirms that ALCA is more structured than SLCA. Conversely, the SLCA selects elements of the matrix with correlation values greater than the ones selected by ALCA and then less affected by statistical uncertainty. 

\section{The Planar Maximally Filtered Graph} \label{PMFGtheory}

The Planar Maximally Filtered Graph has been introduced in a recent paper \cite{Tumminello2005}. The basic idea is to obtain a graph retaining the same hierarchical properties of the MST, i.e. the same hierarchical tree of SLCA, but allowing a greater number of links and more complex topological structures than the MST. The construction of the PMFG is done by relaxing the topological constraint of the MST construction protocol of section \ref{SLCA} according to which no loops are allowed in a tree. Specifically, in the PMFG a link can be included in the graph if and only if the graph with the new link included is still planar. A graph is planar if and only if it can be drawn on a plane without edge crossings \cite{West}. 

The first difference between MST and PMFG is about the number of links, which is $n-1$ in the MST and $3(n-2)$ in the PMFG. Furthermore loops and cliques are allowed in the PMFG. A clique of $r$ elements, r-cliques, is a subgraph of $r$ elements where each element is linked to each other. Because of the Kuratowski's theorem \cite{West} only 3-cliques and 4-cliques are allowed in the PMFG. The study of 3-cliques and 4-cliques is relevant for understanding the strength of clusters in the system \cite{Tumminello2005} as we will see below in the empirical applications. This study can be done through a measure of the intra-cluster connection strength \cite{Tumminello2005}. This measure is obtained by considering a specific sector composed by $n_s$ elements and indicating with $c_4$ and $c_3$ the number of 4-cliques and 3-cliques exclusively composed by elements of the sector. The connection strength $q_s$ of the sector $s$ is therefore defined as
\begin{eqnarray} \label{strength}
          &&    q_s^{4}=\frac{c_4}{n_s-3},\nonumber \\
          &&                       \\      
          &&    q_s^{3}=\frac{c_3}{3 n_s-8}, \nonumber
\end{eqnarray}  
where we distinguish between the connection strength evaluated according to 4-cliques $q_s^{4}$ and 3-cliques $q_s^{3}$. The quantities $n_s-3$ and $3 n_s-8$ are normalizing factors. For large and strongly connected sectors both the measures give almost the same result \cite{Tumminello2005}. When small sectors are considered the quantity $q_s^3$ is more significant than $q_s^4$. Consider for instance a sector of 4 stocks. In this case $q_s^4$ can assume the value 0 or 1, whereas $q_s^3$ can assume one the 5 values 0, 0.25, 0.5, 0.75 and 1, giving a measure of the clustering strength less affected by the quantization error. Note that in this case if $q_s^3$ assumes one of the values 0, 0.25, 0.5 and 0.75 then $q_s^4$ is always zero.
Concerning the hierarchical structure associated to the PMFG it has been shown in Ref. \cite{Tumminello2005} that at any step in the construction of the MST and PMFG, if two elements are connected via at least one path in one of the considered graphs, then they also are connected in the other one. This statement implies that i) the MST is always contained in the PMFG and ii) the hierarchical tree associated to  both the MST and PMFG is the one obtained from the SLCA. 

In summary the PMFG is a graph retaining more information about the system than the MST, the information being stored in the included new links and in the new topological structures allowed. i.e. loops and cliques.

\medskip

\begin{figure}
\begin{center}
              \includegraphics[scale=0.45]{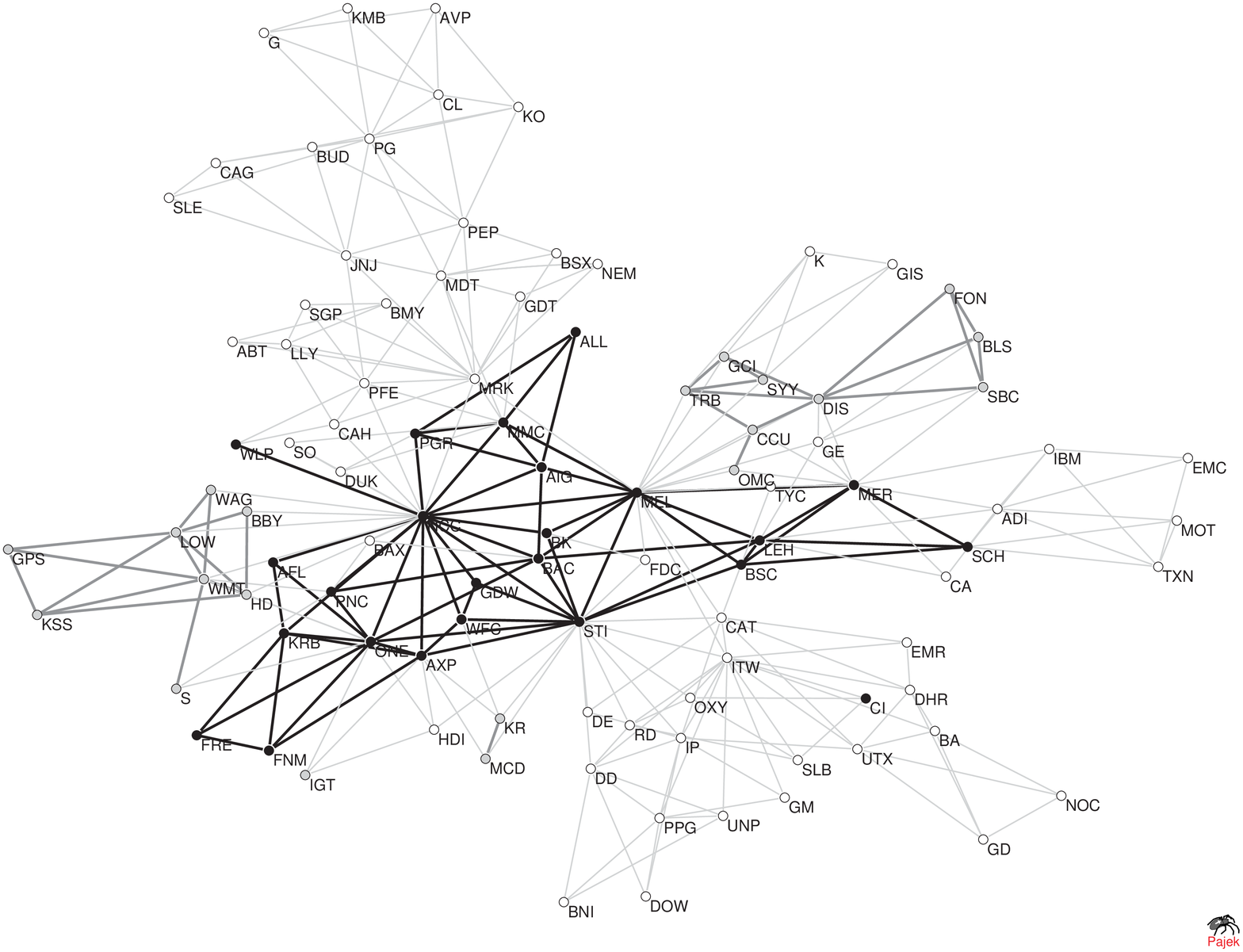}
              \caption{PMFG obtained from daily returns of a set of 100 stocks traded at NYSE in 2002. Black circles are identifying stocks belonging to the Financial sector. Gray circles are identifying stocks belonging to the Services sector. Other stocks are indicated by empty circles. Black thicker lines are connecting stocks belonging to the Financial sector. Gray thicker lines are connecting stocks belonging to the Services sector.}  \label{planarday} 
\end{center}
\end{figure}

The capability of the PMFG of exploiting economic information from a given correlation matrix can be illustrated by considering the set of stocks of Section \ref{dataset}. Here we analyze the topological properties of the PMFG of Fig. \ref{planarday} obtained from the distance matrix of daily returns of the stock set. 

In the figure we again point out the behavior of stocks belonging to the Financial and Services sectors. From the figure we can observe that the Financial sector (black circles) is strongly intra-connected (black thicker edges) whereas for the sector of Services (gray circles) we find just a few intra-sector connections (gray thicker edges). These results agree with the ones observed with the SLCA and the ALCA. The advantage of the study of the PMFG is that, through it, we can perform a quantitative analysis of this behavior, by using the intra--cluster connection strength defined in section \ref{PMFGtheory}.

In Table \ref{tab:PMFG1day} the connection strength is evaluated for all the sectors present in the set. The Financial sector has $q_2^4\approx 0.91$ and $q_2^3\approx 0.91$. This last value is the second biggest value, after the one associated to the Energy. The Energy sector is composed by 3 stocks which are all connected within them so that $q_3^3=1$. The results for the Financial and the Energy sectors agree with the ones shown by both hierarchical trees and the MST. For the Healthcare sector we find $q_6^3\cong 0.68$, thus showing again an high degree of intra-cluster connection. A significant degree of intra-cluster connection can also be observed for the Consumer Non Cyclical sector ($q_3^{4}\cong 0.60$). Both results were also shown by the MST and the hierarchical tree obtained by applying the ALCA. Part of this cluster is also detected by SLCA.  

Among the stocks of the Financial sector, {\em{National City Corp}} (NCC), {\em{Sun Trust Banks Inc}} (STI) and {\em{Mellon Financial Corp}} (MEL) are characterized  by high values of their degree. This fact implies that the Financial sector is strongly connected both within the sector and with other sectors. In particular, NCC is the center of the biggest star in the graph with a degree of 26 (15 are links to stocks of the same sector). On the contrary, the sector of Services is poorly intra-connected: $q_8^{4}=0.18$ and $q_8^{3}\cong 0.31$. This sector is also poorly connected to other sectors. In fact, the highest degree of a stock belonging to this sector is 9 and it is observed for WMT and {\em{Walt Disney}} (DIS). The topology of the PMFG shows that the stocks of the Services sector are mainly gathered in two different subgraphs, clearly visible on the left and the right regions of the graph. The subgraph containing WMT is composed by retail companies. The other one is composed by stocks belonging to subsectors which are different although related the one to each other. In fact, {\em{Sprint Corp FON Group}} (FON), {\em{Bellsouth Corp.}} (BLS) and {\em{SBC Communications}} (SBC) belong to the Communication Services subsector, DIS and {\em{Clear Channel Communications}} (CCU) belong to the Broadcasting subsector, {\em{Tribune Company}} (TRB) and {\em{Gannett Co}} (GCI) belong to the Printing \& Publishing subsector and {\em{Omnicom Group}} (OMC) belongs to the Advertising subsector. The only exception seems to be {\em{Sysco Corp}} (SYY) which belongs to the Retail Grocery subsector. This result provides an example representative of the way in which the PMFG recognizes economic subsectors. The fact that the connection strenghts $q_8^4$ and $q_8^3$ assume low values is indeed explained by the fact that the PMFG detects two subclusters of the same economic sector which are not linked to each other. This is in agreement with the results obtained by using the ALCA and SLCA. In fact, in the case of ALCA we observed two distinct clusters, one of retail companies and another one containing Printing \& Publishing and Broadcasting companies. In the case of the SLCA, which provides a less structured hierarchical tree, only one cluster of retails companies was observed.

In conclusion, we observe that the Financial sector is strongly intra-connected and strongly connected with other sectors. On the other hand, the sector of Services is poorly intra-connected and poorly interacting with other sectors. 
   
\begin{table}\nonumber
\begin{center}
\caption{Intra-sector connection strength (daily returns)} \label{tab:PMFG1day}
\vspace{.5 cm}
\begin{tabular}{||l|c|c|c||}
\hline
SECTOR                & $n_s$ & $q_s^{4}=c_4/[n_s-3]$ & $q_s^{3}=c_3/[3\,n_s-8]$        \\ \hline
Technology            & $ 8$  & $1/5 =0.20$           & $5/16=  0.31$         \\
Financial             & $24$  & $19/21\approx0.91$    & $58/64\approx 0.91$   \\
Energy                & $ 3$  & $--$                  & $1/1=   1.00$         \\
Consumer non-Cyclical & $11$  & $4/8=0.50$            & $15/25\approx 0.60$   \\
Consumer Cyclical     & $ 2$  & $--$                  & $--$                  \\
Healthcare            & $12$  & $5/9\approx0.56$      & $19/28\approx 0.68$   \\
Basic Materials       & $ 6$  & $1/3 \approx0.33$     & $4/10=  0.40$         \\
Services              & $20$  & $3/17\approx0.18$     & $16/52\approx0.31$    \\
Utilities             & $ 2$  & $--$                  & $--$                  \\
Capital Goods         & $ 6$  & $0/3 =0.00$           & $1/10=0.10$           \\
Transportation        & $ 2$  & $--$                  & $--$                  \\
Conglomerates         & $ 4$  & $0/1=0.00$            & $0/4=0.00$            \\ \hline
\end{tabular}
\end{center}
\end{table}

\section{Conclusions}

We observe that all methodologies considered show that the investigated system is structured at daily time horizons. All the methods considered in the present paper are able to detect information about economic sectors of stocks starting from the synchronous correlation coefficient matrix of return time series, although the degree of efficiency in the detection can be different for different techniques. 

In fact, our comparative study shows that the considered methods provide different information about the system. For example, at one day time horizon the method based on RMT predominantly associates the eigenvectors of the highest eigenvalues which are not affected by statistical uncertainty respectively to the market factor (first eigenvalue and eigenvector), the Consumer non-Cyclical sector (second eigenvalue), the Healthcare sector (fourth eigenvalue), the Utilities sector (fifth and nineth eigenvalue) and the Services sector (fifth eigenvalue). In the present case, RMT does not provide information about the existence and strength of economic relation between stocks belonging to the sectors of Technology, Financial, Energy, Consumer Cyclical, Basic Materials, Capital Goods, Transportation and Conglomerates economic sectors. 

A detailed investigation of the hierarchical trees obtained by the SLCA and ALCA shows that these methods are able to efficiently detect various clusters. They can be different from those obtained with the RMT. For example, the Financial sector is well detected by the SLCA and ALCA, while the Technology, the Consumer Non-Cyclical and the Healthcare economic sectors are not well detected at the level of the hierarchical tree. However, it must be mentioned that in the case of the SLCA, the MST can provide distinct yet not independent information about the level of clustering. In particular, the Technology, the Consumer Non-Cyclical and the Healthcare economic sectors are well detected in the MST. The information provided by the MST is usually confirmed by the quantitative analysis of the PMFG performed by computing the intrasector connection strength. 

Our comparative analysis of the hierarchical clustering methods also shows that SLCA is providing information about the highest level of correlation of the correlation matrix whereas the ALCA averages this information within each considered cluster. In this way the average linkage clustering is able to provide a more structured information about the hierarchical organization of the stocks of a set, although it is more affected by statistical uncertainty.

In summary, we believe that our empirical comparison of different methods provides an evidence that RMT and hierarchical clustering methods are able to point out information present in the correlation matrix of the investigated system. The information that is detected with these methods is in part overlapping but in part specific to the selected investigating method. In short, all the approaches detect information but not exactly the same one. For this reason an approach that simultaneously makes use of several of these methods may provide a better characterization of the investigated system than an approach based on just one of them. 

{\bf Acknowledgments} Authors acknowledge support from the research project MIUR-FIRB RBNE01CW3M ``Cellular Self-Organizing nets and chaotic nonlinear dynamics to model and control complex system and from the European Union STREP project n. 012911 ``Human behavior through dynamics of complex social networks: an interdisciplinary approach".


\end{document}